\documentclass[journal=apchd5,manuscript=article]{achemso}

\usepackage{multirow}

\author{Andrzej Szczepkowicz}
\email{andrzej.szczepkowicz@uwr.edu.pl}
\phone{+48 713759304}
\affiliation[University of Wroclaw]
{Institute of Experimental Physics, University of Wroclaw, Plac Maxa Borna 9, 50-204 Wroclaw, Poland}
\author{Po-Wei Kuo}
\author{Yen-Cheh Huang}
\affiliation{Institute of Photonics Technologies, National Tsing Hua University, Hsinchu 300004, Taiwan}

\title[Monochromatic Cherenkov diffraction radiation\ldots]
  {Monochromatic Cherenkov diffraction radiation from continuous coupling of an electron beam to a thin dielectric slab waveguide}


\begin{document}




\begin{abstract}
A scheme for generation of monochromatic Cherenkov radiation in a thin dielectric layer is proposed. The electrons travel in a vacuum parallel to a dielectric, exciting a single synchronous electromagnetic waveguide mode. The proposed scheme is studied quantitatively for near-infrared radiation in silicon induced by a 100-keV electron beam, using time-domain and frequency-domain numerical simulations, with material absorption and dispersion taken into account. This method of radiation generation can be scaled from ultraviolet to terahertz radiation by changing the thickness of the dielectric layer and choosing a material with low loss at the desired wavelength. Comparison with conventional Cherenkov Radiation and Cherenkov Diffraction Radiation is also presented.
\end{abstract}

\section{Introduction}

\begin{figure}
\includegraphics{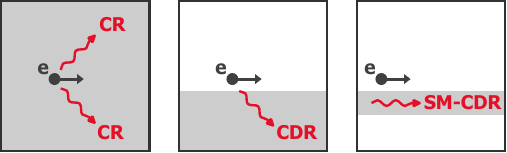}
\caption{Three types of Cherenkov radiation considered in this article.
CR: Cherenkov Radiation (conventional), the electron travels through a uniform dielectric.
CDR: Cherenkov Diffraction Radiation, the electron travels in a vacuum above bulk dielectric.
SM-CDR: Single Mode Cherenkov Diffraction Radiation, the electron travels in a vacuum above a thin dielectric slab.}
\label{fig-intro}
\end{figure}

Cherenkov radiation (CR) is a form of electromagnetic radiation emitted by charged particles traveling with superluminal velocity through a medium \cite{1934-Cherenkov,1937-Cherenkov,1937-Frank-Tamm}. 
CR has numerous well-established applications connected with elementary particle detection \cite{2004-Akimov, 2016-LAnnunziata}. Basic research on CR continues with regard to novel photonic metamaterials 
\cite{
2019-Su-Xiong,
2021-Salas-Montiel,
2023-Gong-Chen,
2023-Roques-Carmes-Kooi, 
2025-Duan}.
CR also continues to be the subject of fundamental physics research \cite{2025-Karlovets-Chaikovskaia}.
For uniform non-dispersive media, CR has a broadband continuous spectrum \cite{1937-Frank-Tamm}. Recently, the possibility of obtaining monochromatic CR was investigated. One possibility is the utilization of dispersion, interference effects, and angular filtering of radiation \cite{2021-Potylitsyn-Kube,2024-Durnic-Potylitsyn}. 
Particle detection based on CR can also be achieved in a non-contact setup, where electrons travel in a vacuum close to the surface of a dielectric medium;  this variant of CR is called ``Cherenkov Diffraction Radiation'' (CDR) \cite{2018-Kieffer-Bartnik, 2020-Curcio-Bergamashi, 2025-Davut-Xia,2021-Konakhovych-Sniezek,2022-Konakhovych-Sniezek}. The difference between conventional CR and CDR is illustrated in Fig.~\ref{fig-intro}. 
Quasi-monochromatic CDR has been experimentally demonstrated with a periodically shaped dielectric target\cite{2022-Karataev-Naumenko}, where an interplay of Cherenkov and Smith‒Purcell radiation can be expected \cite{2021-Konakhovych-Sniezek,2022-Konakhovych-Sniezek, 2023-Szczepkowicz-Konakhovych}. A different route for monochromatization of CDR is the pre-bunching of the electron beam for coherent electromagnetic emission \cite{2020-Karataev-Fedorov}. Another scheme for monochromatic CDR was demonstrated in the context of generation of high-power terahertz radiation, where an electron beam travels inside a hollow cylindrical dielectric tube surrounded by a metallic wall (dielectric-lined waveguide) and couples to a discrete set of waveguide modes \cite{2009-Cook-Tikhoplav}. The physical principle of this scheme is closest to the setup proposed in the present work, but differs by the presence of a metallic wall, and by topology: cylindrical vs.\ planar.  

In the present work we propose a simple, scalable scheme for monochromatization of CDR, which is compatible with planar on-chip photonic nanofabrication technology and does not require beam bunching. 
The proposed scheme is connected with recent research on miniature, laser-driven accelerators \cite{2021-England-Hommelhoff-Byer,2022-Shiloh-Schonenberger}, because in principle it is an inversion of one of the proposed electron acceleration schemes, where the accelerating laser light travels in a dielectric waveguide along the electron beam \cite{2017-Kozak-Beck, 2018-Zhao-Hughes, 2023-Palmeri-Salerno}.
In our proposed radiation scheme the electron travels in a vacuum, in proximity of a thin dielectric slab waveguide, as shown in Fig.~\ref{fig-intro}(c). The resulting radiation is guided by the dielectric layer, and, as shown in Sect.~\ref{results-discussion}, ultimately only one waveguide mode survives. We call the resulting radiation Single-Mode Cherenkov Diffraction Radiation (SM-CDR). Common to all the three forms of Cherenkov radiation shown in Fig.~\ref{fig-intro} is the requirement that the particle velocity is above the Cherenkov threshold, $v>c/n$, where $n$ is the index of refraction of the dielectric medium. For the case of SM-CDR, the fulfillment of this condition is illustrated in the dispersion diagram shown in the next section (Fig.~\ref{fig-waveguide-dispersion}).

Interaction of free electrons with a dielectric slab has been reported in the literature for electrons penetrating the dielectric sample \cite{1975-Chen-Silcox,2004-de-Abajo-Rivacoba,2010-Couillard-Yurtsever,2013-Saito-Chen,2025-Preimesberger-Hornof}. The present setup, with electrons moving in a vacuum along the slab waveguide, could be treated analytically, either by considering multiple successive reflections of plane-wave components at both inner boundaries\cite{2004-de-Abajo-Rivacoba}, or by the polarization current approach\cite{2011-Karlovets, 2015-Shevelev-Konkov}. However, to the best of our knowledge, no theoretical or experimental guided radiation spectra have so far been published for the setup proposed here.

The proposed mechanism of generating monochromatic radiation scheme is general and can be scaled from ultraviolet to terahertz radiation by changing the thickness of the dielectric layer and choosing a material with low loss at the desired wavelength. In this paper we focus on one specific example: generation of near-infrared radiation in a silicon slab of thickness $0.2~\mu$m -- much thinner than the target wavelength ($1.4~\mu$m). We choose to demonstrate the phenomenon on a simplest, high symmetry model, so we assume that the slab is surrounded by vacuum, which also facilitates analytical calculation of the dispersion structure\cite{1999-Inan-Inan}. In a real experimental configuration, the Si slab would form a layer on some low index material like SiO$_2$. This would slightly change the eigenmode frequencies, but the essential physics would remain the same.

\section{\label{sect-methods}Methods}
In this work we consider a 100 keV electron (velocity $v_e=0.5482c$) interacting with a $0.2~\mu$m thick silicon slab at a distance (impact parameter) of $0.1~\mu$m. 
We take into account material dispersion and absorption \cite{2024-refractiveindex-info}. The refractive index changes from $n=3.43$ at 21~THz to $n=5.11$ at 720~THz; the corresponding Cherenkov angles change from $58^{\circ}$ to $69^{\circ}$. The material transparency window lies in the near infrared; in the frequency range 64--300~THz the extinction coefficient $k$ is below 0.001 \cite{2019-Saleh-Teich,2024-refractiveindex-info}.

\begin{figure}
\includegraphics{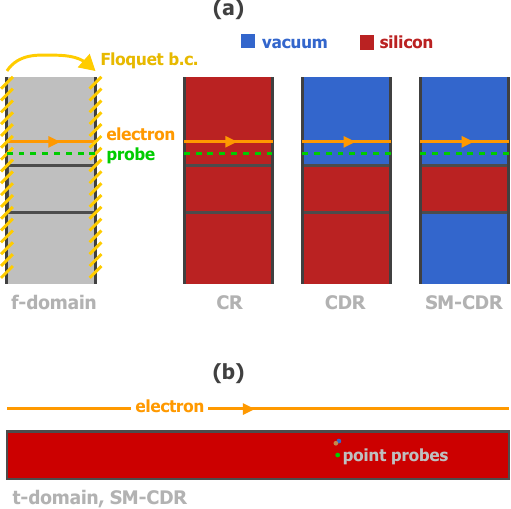}
\caption{Geometry of the numerical models. (a) The frequency-domain model with the Floquet boundary condition shown in yellow and numerical probe surface (for Poynting vector integration) indicated by a green line. (b) The time-domain model.}
\label{fig-methods}
\end{figure}

The numerical results are obtained using a combination of commercial solvers (CST Studio Suite and Comsol Multiphysics): three-dimensional finite-difference time-domain models, three-dimensional and two-dimensional time-domain and frequency-domain finite-element models \cite{2020-Szczepkowicz-Schachter-England, 2021-Konakhovych-Sniezek, 2022-Konakhovych-Sniezek}. The geometry of the models is shown in Fig.~\ref{fig-methods}. For SM-CDR, the absolute rate of energy transfer from the electron to the waveguide is computed in the time-domain model by integrating the Poynting vector over the surface of the waveguide facing the electron, and checked by calculating the rate of electromagnetic energy increase inside the waveguide; the spectrum of the guided radiation is obtained by Fourier transforming the signal from the point probes shown in Fig.~\ref{fig-methods}(b). The relative rates of energy transfer for CR, CDR, and SM-CDR, are computed by integrating the Poynting vector along the probe surfaces shown in Fig.~\ref{fig-methods}(a).

\section{\label{results-discussion}Results and discussion} 

\subsection{\label{sect-waveguide-dispersion}Waveguide modes with material dispersion} 

\begin{figure}
\includegraphics{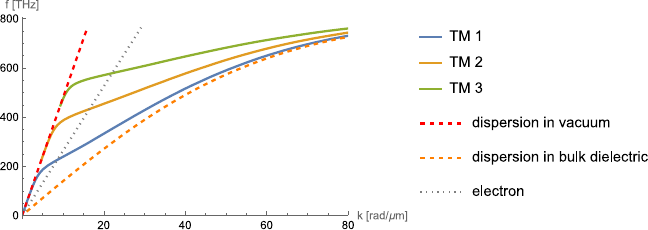}
\caption{Analytical waveguide dispersion curves of the 0.2~$\mu$m thick silicon slab waveguide, with material dispersion taken into account.}
\label{fig-waveguide-dispersion}
\end{figure}

An electron beam traveling in the vicinity of a dielectric slab waveguide couples to transverse-magnetic waveguide modes (TM modes). In the first step we solve transcendental equations for the guided wave\cite{1999-Inan-Inan} to calculate waveguide dispersion curves for TM modes, for a silicon slab of thickness $a=0.2~\mu$m, with material dispersion, $n=n(f)$, taken into account (see also Refs.\cite{1975-Chen-Silcox,2025-Preimesberger-Hornof}). Figure \ref{fig-waveguide-dispersion} shows the dispersion curves for the first three TM modes, together with lines corresponding to characteristic velocities: speed of light $c$, speed of light in bulk dielectric $c/n(f)$, and electron velocity  $v_e=0.5482$ (for 100~keV electrons). The electron line intersects 
TM1 at $f_1=225$~THz, 
TM2 at $f_2=432$~THz, and 
TM3 at $f_3=578$~THz etc. 
This means that initially several eigenmodes are excited by the electron. However, only $f_1$ lies within the transparency window of silicon \cite{2019-Saleh-Teich}, so higher order modes are attenuated and only monochromatic radiation at $f_1=225$~THz (corresponding to vacuum wavelength $\lambda=1.33~\mu$m) will survive as the wave train will propagate in the silicon waveguide.

\subsection{\label{sect-comparison}Comparison of three forms of Cherenkov radiation: CR, CDR, SM-CDR}

\begin{figure}
\includegraphics{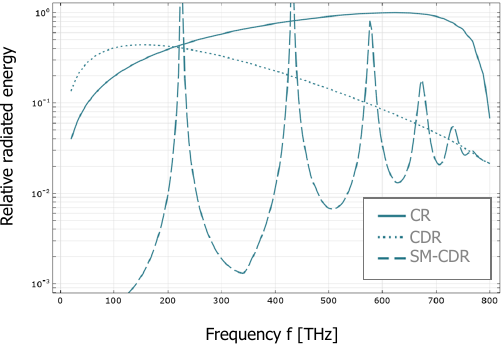}
\caption{Comparison of three radiation spectra representing the three types of Cherenkov radiation, obtained within a frequency-domain numerical model. The maxima for SM-CDR occur at frequencies: 225, 432, 578 THz.}
\label{fig-three-spectra}
\end{figure}

In Fig.~\ref{fig-three-spectra} we compare the relative radiation spectra of CR, CDR and SM-CDR for the three corresponding geometries. The simulation compares the energy flow from the electron to the dielectric medium (Poynting vector integrated over the indicated ``probe surface'') in a two-dimensional frequency-domain simulation \cite{2020-Szczepkowicz-Schachter-England}.

For all three types of radiation, we observe a cutoff at high frequencies due to absorption in silicon. According to Frank-Tamm theory \cite{1937-Frank-Tamm}, assuming constant $n$ and no absorption, the CR spectrum should be simply proportional to $f$. The CR spectrum in Fig.~\ref{fig-three-spectra} shows a more complex behavior due to material dispersion and absorption.
The diffraction radiation variant (CDR) is expected to differ from conventional CR by an additional multiplyer 
$e^{-4\pi f y / c \beta \gamma}$ 
($y$ = impact parameter, $\beta,\gamma$ 
-- relativistic parameters of the electron beam)\cite{2010-Potylitsyn-Ryazanov}. This exponential decay with $f$ is consistent with the CDR curve in Fig.~\ref{fig-three-spectra}. The third curve in the figure, the SM-CDR spectrum, is one of the main numerical results of the present study. In the literature one cannot find the theory for this spectrum. The resonant frequencies can be predicted on the basis of the theoretical dispersion diagram shown in Fig.~\ref{fig-waveguide-dispersion} -- the analytical and numerical frequencies $f_1$, $f_2$ and $f_3$ agree with accuracy better then 1~THz. Although this SM-CDR spectrum appears multimodal as studied close to the location of the electron (see the location of probe surface in Fig.~\ref{fig-methods}(a)), only the first peak is within the transparency window of silicon, and as the wave is guided in the silicon slab, only a single guided mode survives.

\subsection{\label{sect-wavetrain}Generation and evolution of the SM-CDR wave train}

\begin{figure}
\includegraphics{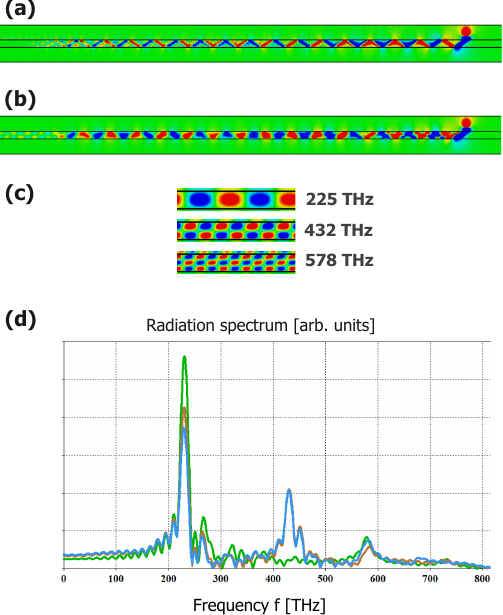}
\caption{SM-CDR radiation wave train generated inside the silicon slab after 100~fs of electron-dielectric interaction, obtained within a time-domain numerical model. (a)~and (b) show the transverse magnetic field of the travelling electron (right hand side) and the generated guided radiation. 
(a)~Model without absorption and dispersion. 
(b)~Model with absorption and dispersion. Here the guided radiation gradually evolves into a single guided mode (TM1), as the higher frequencies are outside the transparency window of silicon. 
(c)~The result of frequency filtering applied to the field pattern in (b). 
(d)~The guided radiation spectrum corresponding to~(b), taken at different depths below the top surface of the slab to show different waveguide modes (see Fig.~\ref{fig-methods}(b) for point probe positions).}
\label{fig-wavetrain}
\end{figure}

The generation of the guided electromagnetic wave train begins when the electron appears above the dielectric slab, and the length of the Cherenkov wave train increases in time. Figure~\ref{fig-wavetrain} shows the wave train after 100~fs of interaction. 

The energy is transferred from the electron to the waveguide at a rate of $0.16~\mu$W (per electron, at impact parameter $0.1~\mu$m), as determined from the time-domain numerical model. When material absorption is taken into account, the wave train gradually evolves from a superposition of modes towards a single mode pattern.
The length of the wave train is proportional to the time of interaction between the electron and the slab: $l = (v_{\mathrm{head}} - v_{\mathrm{tail}}) t$. Note that the Cherenkov wave train is not a free-traveling wave packet. 
On the one hand, the wave train is continuously generated at its head, so $v_{\mathrm{head}}$ is equal to the electron velocity ($v_e=0.5482c$) and to the phase velocities of the constituent waveguide modes excited by the electron (see also Fig.~\ref{fig-waveguide-dispersion}). 
On the other hand, the tail of the wave train is subject to free propagation in the dielectric slab waveguide. Its numerically measured velocity is $v_{\mathrm{tail}}=0.1889(48)$. This is close to the theoretical phase velocity of the TM1 mode ($0.193c$), as calculated from the slope of the dispersion curve at the intersection with the electron line, see Fig.~\ref{fig-waveguide-dispersion}. The first mode is the most relevant, as ultimately it is the only mode that survives due to electromagnetic absorption in silicon. For completeness, the phase velocities of the first three TM modes are $0.193c$, $0.130c$ and $0.079c$. 

Note that when the guided wave train evolves into true Single Mode Chrenkov Diffraction Radiation (SM-CDR), the Cherenkov wavefront (or the ``shock wave'') is no longer present, because such a shock wave is polychromatic by nature. It can still be seen in the numerical model if material absorption is neglected (Fig.~\ref{fig-wavetrain}(a)), but, in the model with absorption (Fig.~\ref{fig-wavetrain}(b)), the Cherenkov wavefront disappears as the higher frequency components are absorbed in silicon.

\section{Conclusion}
While conventional Cherenkov Radiation (CR) has a broad spectrum, this work shows a very simple way to obtain monochromatic CR, without advanced materials, complex structures or electron beam bunching. The phenomenon was quantitatively demonstrated numerically in a combination of time-domain and frequency-domain models. In the proposed scheme the electrons travel in a vacuum in the vicinity of a thin dielectric slab waveguide. The electrons couple to a few discrete waveguide modes, but only one mode lies in the transparency window of silicon and survives the propagation of the wave train in the waveguide. 
The head of the Cherenkov wave train moves with phase velocity of the excited electromagnetic mode, which is synchronous with the electron, while the tail propagates freely with the group velocity of the mode -- the difference in head and tail velocities account for the rate of growth of the Cherenkov wave train.
The setup presented here is scalable from terahertz to ultraviolet radiation, if appropriate material and waveguide thickness is chosen.
The peculiar form of CR proposed here might have impact on traditional domain of application of CR: elementary particle detection and beam diagnostics\cite{2025-Davut-Xia}. It may also be useful for recent investigations on quantum interaction of electron beams with light\cite{2023-Roques-Carmes-Kooi}, because with monochromatic light one can expect the depletion of electron kinetic energy in well defined quanta.

\begin{acknowledgement}
We gratefully acknowledge discussions with Joel England, Levi Schachter, Yen-Hung Chen, Hossein Shivrani, Alexey Kopeykin, Timofey Panfilov and Dmitry Karlovets.
We gratefully acknowledge that the CST codes created in this work were in part based on previous work by Dmytro Konakhovych \cite{2021-Konakhovych-Sniezek}.

\noindent \textbf{Funding:}\\
Po-Wei Kuo and Yen-Chieh Huang acknowledge funding support from National Science and
Technology Council of Taiwan under Contracts NSTC 113-2923-M-007-001, NSTC 112-2112-M-007 -028 -MY3.
\end{acknowledgement}

\section*{Disclosures}
The authors declare no conflicts of interest.

\bibliography{article-si-slab.bib}

\providecommand{\latin}[1]{#1}
\makeatletter
\providecommand{\doi}
  {\begingroup\let\do\@makeother\dospecials
  \catcode`\{=1 \catcode`\}=2 \doi@aux}
\providecommand{\doi@aux}[1]{\endgroup\texttt{#1}}
\makeatother
\providecommand*\mcitethebibliography{\thebibliography}
\csname @ifundefined\endcsname{endmcitethebibliography}  {\let\endmcitethebibliography\endthebibliography}{}
\begin{mcitethebibliography}{40}
\providecommand*\natexlab[1]{#1}
\providecommand*\mciteSetBstSublistMode[1]{}
\providecommand*\mciteSetBstMaxWidthForm[2]{}
\providecommand*\mciteBstWouldAddEndPuncttrue
  {\def\EndOfBibitem{\unskip.}}
\providecommand*\mciteBstWouldAddEndPunctfalse
  {\let\EndOfBibitem\relax}
\providecommand*\mciteSetBstMidEndSepPunct[3]{}
\providecommand*\mciteSetBstSublistLabelBeginEnd[3]{}
\providecommand*\EndOfBibitem{}
\mciteSetBstSublistMode{f}
\mciteSetBstMaxWidthForm{subitem}{(\alph{mcitesubitemcount})}
\mciteSetBstSublistLabelBeginEnd
  {\mcitemaxwidthsubitemform\space}
  {\relax}
  {\relax}

\bibitem[{\v{C}}erenkov(1934)]{1934-Cherenkov}
{\v{C}}erenkov,~P. Visible light from pure liquids under the impact of gamma-rays. \emph{Comptes rendus de l'Acad\'{e}mie des sciences de l'{URSS}} \textbf{1934}, \emph{3}, 451--457\relax
\mciteBstWouldAddEndPuncttrue
\mciteSetBstMidEndSepPunct{\mcitedefaultmidpunct}
{\mcitedefaultendpunct}{\mcitedefaultseppunct}\relax
\EndOfBibitem
\bibitem[{\v{C}}erenkov(1937)]{1937-Cherenkov}
{\v{C}}erenkov,~P.~A. Visible Radiation Produced by Electrons Moving in a Medium with Velocities Exceeding that of Light. \emph{Physical Review} \textbf{1937}, \emph{52}, 378--379\relax
\mciteBstWouldAddEndPuncttrue
\mciteSetBstMidEndSepPunct{\mcitedefaultmidpunct}
{\mcitedefaultendpunct}{\mcitedefaultseppunct}\relax
\EndOfBibitem
\bibitem[Frank and Tamm(1937)Frank, and Tamm]{1937-Frank-Tamm}
Frank,~I.; Tamm,~I. Coherent visible radiation of fast electrons passing through matter. \emph{Comptes rendus de l'Acad\'{e}mie des sciences de l'{URSS}} \textbf{1937}, \emph{14}, 109--114\relax
\mciteBstWouldAddEndPuncttrue
\mciteSetBstMidEndSepPunct{\mcitedefaultmidpunct}
{\mcitedefaultendpunct}{\mcitedefaultseppunct}\relax
\EndOfBibitem
\bibitem[Akimov(2004)]{2004-Akimov}
Akimov,~Y.~K. Cherenkov detectors in particle physics. \emph{Physics of Atomic Nuclei} \textbf{2004}, \emph{67}, 1385--1389\relax
\mciteBstWouldAddEndPuncttrue
\mciteSetBstMidEndSepPunct{\mcitedefaultmidpunct}
{\mcitedefaultendpunct}{\mcitedefaultseppunct}\relax
\EndOfBibitem
\bibitem[L’Annunziata(2016)]{2016-LAnnunziata}
L’Annunziata,~M.~F. \emph{Radioactivity}; Elsevier, 2016; p 547–581\relax
\mciteBstWouldAddEndPuncttrue
\mciteSetBstMidEndSepPunct{\mcitedefaultmidpunct}
{\mcitedefaultendpunct}{\mcitedefaultseppunct}\relax
\EndOfBibitem
\bibitem[Su \latin{et~al.}(2019)Su, Xiong, Xu, Cai, Yin, Peng, and Liu]{2019-Su-Xiong}
Su,~Z.; Xiong,~B.; Xu,~Y.; Cai,~Z.; Yin,~J.; Peng,~R.; Liu,~Y. Manipulating Cherenkov Radiation and Smith–Purcell Radiation by Artificial Structures. \emph{Advanced Optical Materials} \textbf{2019}, \emph{7}, 1801666\relax
\mciteBstWouldAddEndPuncttrue
\mciteSetBstMidEndSepPunct{\mcitedefaultmidpunct}
{\mcitedefaultendpunct}{\mcitedefaultseppunct}\relax
\EndOfBibitem
\bibitem[Salas-Montiel(2021)]{2021-Salas-Montiel}
Salas-Montiel,~R. Cherenkov radiation in integrated nanophotonic structures. \emph{Journal of Applied Physics} \textbf{2021}, \emph{129}, 233103\relax
\mciteBstWouldAddEndPuncttrue
\mciteSetBstMidEndSepPunct{\mcitedefaultmidpunct}
{\mcitedefaultendpunct}{\mcitedefaultseppunct}\relax
\EndOfBibitem
\bibitem[Gong \latin{et~al.}(2023)Gong, Chen, Chen, Zhu, Wang, Zhang, Hu, Yang, Zhang, Chen, Kaminer, and Lin]{2023-Gong-Chen}
Gong,~Z.; Chen,~J.; Chen,~R.; Zhu,~X.; Wang,~C.; Zhang,~X.; Hu,~H.; Yang,~Y.; Zhang,~B.; Chen,~H.; Kaminer,~I.; Lin,~X. Interfacial Cherenkov radiation from ultralow-energy electrons. \emph{Proceedings of the National Academy of Sciences} \textbf{2023}, \emph{120}, e2306601120\relax
\mciteBstWouldAddEndPuncttrue
\mciteSetBstMidEndSepPunct{\mcitedefaultmidpunct}
{\mcitedefaultendpunct}{\mcitedefaultseppunct}\relax
\EndOfBibitem
\bibitem[Roques-Carmes \latin{et~al.}(2023)Roques-Carmes, Kooi, Yang, Rivera, Keathley, Joannopoulos, Johnson, Kaminer, Berggren, and Solja{\v{c}}i{\'{c}}]{2023-Roques-Carmes-Kooi}
Roques-Carmes,~C.; Kooi,~S.~E.; Yang,~Y.; Rivera,~N.; Keathley,~P.~D.; Joannopoulos,~J.~D.; Johnson,~S.~G.; Kaminer,~I.; Berggren,~K.~K.; Solja{\v{c}}i{\'{c}},~M. Free-electron{\textendash}light interactions in nanophotonics. \emph{Applied Physics Reviews} \textbf{2023}, \emph{10}, 011303\relax
\mciteBstWouldAddEndPuncttrue
\mciteSetBstMidEndSepPunct{\mcitedefaultmidpunct}
{\mcitedefaultendpunct}{\mcitedefaultseppunct}\relax
\EndOfBibitem
\bibitem[Duan(2025)]{2025-Duan}
Duan,~Z. \emph{Metamaterial-Based Electromagnetic Radiations and Applications}; Springer Nature Singapore, 2025\relax
\mciteBstWouldAddEndPuncttrue
\mciteSetBstMidEndSepPunct{\mcitedefaultmidpunct}
{\mcitedefaultendpunct}{\mcitedefaultseppunct}\relax
\EndOfBibitem
\bibitem[Karlovets \latin{et~al.}(2025)Karlovets, Chaikovskaia, Grosman, Kargina, Shchepkin, and Sizykh]{2025-Karlovets-Chaikovskaia}
Karlovets,~D.; Chaikovskaia,~A.; Grosman,~D.; Kargina,~D.; Shchepkin,~A.; Sizykh,~G. Attosecond physics hidden in Cherenkov radiation. \emph{Communications Physics} \textbf{2025}, \emph{8}, Article 192\relax
\mciteBstWouldAddEndPuncttrue
\mciteSetBstMidEndSepPunct{\mcitedefaultmidpunct}
{\mcitedefaultendpunct}{\mcitedefaultseppunct}\relax
\EndOfBibitem
\bibitem[Potylitsyn \latin{et~al.}(2021)Potylitsyn, Kube, Novokshonov, Vukolov, Gogolev, Alexeev, Klag, and Lauth]{2021-Potylitsyn-Kube}
Potylitsyn,~A.; Kube,~G.; Novokshonov,~A.; Vukolov,~A.; Gogolev,~S.; Alexeev,~B.; Klag,~P.; Lauth,~W. First observation of quasi–monochromatic optical Cherenkov radiation in a dispersive medium (quartz). \emph{Physics Letters A} \textbf{2021}, \emph{417}, 127680\relax
\mciteBstWouldAddEndPuncttrue
\mciteSetBstMidEndSepPunct{\mcitedefaultmidpunct}
{\mcitedefaultendpunct}{\mcitedefaultseppunct}\relax
\EndOfBibitem
\bibitem[Durnic \latin{et~al.}(2024)Durnic, Potylitsyn, Bogdanov, and Gogolev]{2024-Durnic-Potylitsyn}
Durnic,~B.; Potylitsyn,~A.; Bogdanov,~A.; Gogolev,~S. Radiator thickness and its effects on Cherenkov spectral lines. \emph{Nuclear Instruments and Methods in Physics Research Section A: Accelerators, Spectrometers, Detectors and Associated Equipment} \textbf{2024}, \emph{1059}, 169015\relax
\mciteBstWouldAddEndPuncttrue
\mciteSetBstMidEndSepPunct{\mcitedefaultmidpunct}
{\mcitedefaultendpunct}{\mcitedefaultseppunct}\relax
\EndOfBibitem
\bibitem[Kieffer \latin{et~al.}(2018)Kieffer, Bartnik, Bergamaschi, Bleko, Billing, Bobb, Conway, Forster, Karataev, Konkov, Jones, Lefevre, Markova, Mazzoni, Padilla~Fuentes, Potylitsyn, Shanks, and Wang]{2018-Kieffer-Bartnik}
Kieffer,~R. \latin{et~al.}  Direct Observation of Incoherent Cherenkov Diffraction Radiation in the Visible Range. \emph{Phys. Rev. Lett.} \textbf{2018}, \emph{121}, 054802\relax
\mciteBstWouldAddEndPuncttrue
\mciteSetBstMidEndSepPunct{\mcitedefaultmidpunct}
{\mcitedefaultendpunct}{\mcitedefaultseppunct}\relax
\EndOfBibitem
\bibitem[Curcio \latin{et~al.}(2020)Curcio, Bergamaschi, Corsini, Farabolini, Gamba, Garolfi, Kieffer, Lefevre, Mazzoni, Fedorov, Gardelle, Gilardi, Karataev, Lekomtsev, Pacey, Saveliev, Potylitsyn, and Senes]{2020-Curcio-Bergamashi}
Curcio,~A. \latin{et~al.}  Noninvasive bunch length measurements exploiting Cherenkov diffraction radiation. \emph{Phys. Rev. Accel. Beams} \textbf{2020}, \emph{23}, 022802\relax
\mciteBstWouldAddEndPuncttrue
\mciteSetBstMidEndSepPunct{\mcitedefaultmidpunct}
{\mcitedefaultendpunct}{\mcitedefaultseppunct}\relax
\EndOfBibitem
\bibitem[Davut \latin{et~al.}(2025)Davut, Xia, Apsimon, McGunigal, Karataev, Lefevre, Mazzoni, and Senes]{2025-Davut-Xia}
Davut,~C.; Xia,~G.; Apsimon,~O.; McGunigal,~J.; Karataev,~P.; Lefevre,~T.; Mazzoni,~S.; Senes,~E. Design and experimental verification of a bunch length monitor based on coherent Cherenkov diffraction radiation. \emph{Physical Review Research} \textbf{2025}, \emph{7}, 013193\relax
\mciteBstWouldAddEndPuncttrue
\mciteSetBstMidEndSepPunct{\mcitedefaultmidpunct}
{\mcitedefaultendpunct}{\mcitedefaultseppunct}\relax
\EndOfBibitem
\bibitem[Konakhovych \latin{et~al.}(2021)Konakhovych, Sniezek, Warmusz, Black, Zhao, England, and Szczepkowicz]{2021-Konakhovych-Sniezek}
Konakhovych,~D.; Sniezek,~D.; Warmusz,~O.; Black,~D.~S.; Zhao,~Z.; England,~R.~J.; Szczepkowicz,~A. Internal Smith-Purcell radiation and its interplay with Cherenkov diffraction radiation in silicon -- a combined time and frequency domain numerical study. 2021; \url{https://arxiv.org/abs/2105.07682}, arXiv:2105.07682\relax
\mciteBstWouldAddEndPuncttrue
\mciteSetBstMidEndSepPunct{\mcitedefaultmidpunct}
{\mcitedefaultendpunct}{\mcitedefaultseppunct}\relax
\EndOfBibitem
\bibitem[Konakhovych \latin{et~al.}(2022)Konakhovych, Sniezek, Warmusz, Black, Zhao, England, Tishchenko, Sergeeva, Ponomarenko, and Szczepkowicz]{2022-Konakhovych-Sniezek}
Konakhovych,~D.; Sniezek,~D.; Warmusz,~O.; Black,~D.~S.; Zhao,~Z.; England,~R.~J.; Tishchenko,~A.~A.; Sergeeva,~D.~Y.; Ponomarenko,~A.~A.; Szczepkowicz,~A. Internal Smith--Purcell radiation and its interplay with Cherenkov diffraction radiation in silicon -- a combined time and frequency domain study. 2022; unpublished manuscript\relax
\mciteBstWouldAddEndPuncttrue
\mciteSetBstMidEndSepPunct{\mcitedefaultmidpunct}
{\mcitedefaultendpunct}{\mcitedefaultseppunct}\relax
\EndOfBibitem
\bibitem[Karataev \latin{et~al.}(2022)Karataev, Naumenko, Potylitsyn, Shevelev, and Artyomov]{2022-Karataev-Naumenko}
Karataev,~P.; Naumenko,~G.; Potylitsyn,~A.; Shevelev,~M.; Artyomov,~K. Observation of quasi-monochromatic resonant Cherenkov diffraction radiation. \emph{Results in Physics} \textbf{2022}, \emph{33}, 105079\relax
\mciteBstWouldAddEndPuncttrue
\mciteSetBstMidEndSepPunct{\mcitedefaultmidpunct}
{\mcitedefaultendpunct}{\mcitedefaultseppunct}\relax
\EndOfBibitem
\bibitem[Szczepkowicz \latin{et~al.}(2023)Szczepkowicz, Konakhovych, Sniezek, Black, England, Huang, and Schachter]{2023-Szczepkowicz-Konakhovych}
Szczepkowicz,~A.; Konakhovych,~D.; Sniezek,~D.; Black,~D.~S.; England,~R.~J.; Huang,~Y.-C.; Schachter,~L. Efficiency of gratings for silica fiber-coupled internal Smith-Purcell radiation and Cherenkov diffraction radiation -- a quantitative numerical study. 2023; \url{https://arxiv.org/abs/2310.20520}, arXiv:2310.20520\relax
\mciteBstWouldAddEndPuncttrue
\mciteSetBstMidEndSepPunct{\mcitedefaultmidpunct}
{\mcitedefaultendpunct}{\mcitedefaultseppunct}\relax
\EndOfBibitem
\bibitem[Karataev \latin{et~al.}(2020)Karataev, Fedorov, Naumenko, Popov, Potylitsyn, and Vukolov]{2020-Karataev-Fedorov}
Karataev,~P.; Fedorov,~K.; Naumenko,~G.; Popov,~K.; Potylitsyn,~A.; Vukolov,~A. Ultra-monochromatic far-infrared Cherenkov diffraction radiation in a super-radiant regime. \emph{Scientific Reports} \textbf{2020}, \emph{10}, Article 20961\relax
\mciteBstWouldAddEndPuncttrue
\mciteSetBstMidEndSepPunct{\mcitedefaultmidpunct}
{\mcitedefaultendpunct}{\mcitedefaultseppunct}\relax
\EndOfBibitem
\bibitem[Cook \latin{et~al.}(2009)Cook, Tikhoplav, Tochitsky, Travish, Williams, and Rosenzweig]{2009-Cook-Tikhoplav}
Cook,~A.~M.; Tikhoplav,~R.; Tochitsky,~S.~Y.; Travish,~G.; Williams,~O.~B.; Rosenzweig,~J.~B. Observation of Narrow-Band Terahertz Coherent Cherenkov Radiation from a Cylindrical Dielectric-Lined Waveguide. \emph{Physical Review Letters} \textbf{2009}, \emph{103}, 095003\relax
\mciteBstWouldAddEndPuncttrue
\mciteSetBstMidEndSepPunct{\mcitedefaultmidpunct}
{\mcitedefaultendpunct}{\mcitedefaultseppunct}\relax
\EndOfBibitem
\bibitem[England \latin{et~al.}(2021)England, Hommelhoff, and Byer]{2021-England-Hommelhoff-Byer}
England,~R.~J.; Hommelhoff,~P.; Byer,~R.~L. Microchip accelerators. \emph{Physics Today} \textbf{2021}, \emph{74}, 42–49\relax
\mciteBstWouldAddEndPuncttrue
\mciteSetBstMidEndSepPunct{\mcitedefaultmidpunct}
{\mcitedefaultendpunct}{\mcitedefaultseppunct}\relax
\EndOfBibitem
\bibitem[Shiloh \latin{et~al.}(2022)Shiloh, Sch\"{o}nenberger, Adiv, Ruimy, Karnieli, Hughes, England, Leedle, Black, Zhao, Musumeci, Byer, Arie, Kaminer, and Hommelhoff]{2022-Shiloh-Schonenberger}
Shiloh,~R.; Sch\"{o}nenberger,~N.; Adiv,~Y.; Ruimy,~R.; Karnieli,~A.; Hughes,~T.; England,~R.~J.; Leedle,~K.~J.; Black,~D.~S.; Zhao,~Z.; Musumeci,~P.; Byer,~R.~L.; Arie,~A.; Kaminer,~I.; Hommelhoff,~P. Miniature light-driven nanophotonic electron acceleration and control. \emph{Advances in Optics and Photonics} \textbf{2022}, \emph{14}, 862\relax
\mciteBstWouldAddEndPuncttrue
\mciteSetBstMidEndSepPunct{\mcitedefaultmidpunct}
{\mcitedefaultendpunct}{\mcitedefaultseppunct}\relax
\EndOfBibitem
\bibitem[Kozák \latin{et~al.}(2017)Kozák, Beck, Deng, McNeur, Sch\"{o}nenberger, Gaida, Stutzki, Gebhardt, Limpert, Ruehl, Hartl, Solgaard, Harris, Byer, and Hommelhoff]{2017-Kozak-Beck}
Kozák,~M.; Beck,~P.; Deng,~H.; McNeur,~J.; Sch\"{o}nenberger,~N.; Gaida,~C.; Stutzki,~F.; Gebhardt,~M.; Limpert,~J.; Ruehl,~A.; Hartl,~I.; Solgaard,~O.; Harris,~J.~S.; Byer,~R.~L.; Hommelhoff,~P. Acceleration of sub-relativistic electrons with an evanescent optical wave at a planar interface. \emph{Optics Express} \textbf{2017}, \emph{25}, 19195\relax
\mciteBstWouldAddEndPuncttrue
\mciteSetBstMidEndSepPunct{\mcitedefaultmidpunct}
{\mcitedefaultendpunct}{\mcitedefaultseppunct}\relax
\EndOfBibitem
\bibitem[Zhao \latin{et~al.}(2018)Zhao, Hughes, Tan, Deng, Sapra, England, Vuckovic, Harris, Byer, and Fan]{2018-Zhao-Hughes}
Zhao,~Z.; Hughes,~T.~W.; Tan,~S.; Deng,~H.; Sapra,~N.; England,~R.~J.; Vuckovic,~J.; Harris,~J.~S.; Byer,~R.~L.; Fan,~S. Design of a tapered slot waveguide dielectric laser accelerator for sub-relativistic electrons. \emph{Optics Express} \textbf{2018}, \emph{26}, 22801\relax
\mciteBstWouldAddEndPuncttrue
\mciteSetBstMidEndSepPunct{\mcitedefaultmidpunct}
{\mcitedefaultendpunct}{\mcitedefaultseppunct}\relax
\EndOfBibitem
\bibitem[Palmeri \latin{et~al.}(2023)Palmeri, Salerno, Mauro, Rocco, Locatelli, Torrisi, and Sorbello]{2023-Palmeri-Salerno}
Palmeri,~R.; Salerno,~N.; Mauro,~G.~S.; Rocco,~D.; Locatelli,~A.; Torrisi,~G.; Sorbello,~G. Optimization of sub-relativistic co-propagating accelerating structures. \emph{Optics Express} \textbf{2023}, \emph{31}, 38891\relax
\mciteBstWouldAddEndPuncttrue
\mciteSetBstMidEndSepPunct{\mcitedefaultmidpunct}
{\mcitedefaultendpunct}{\mcitedefaultseppunct}\relax
\EndOfBibitem
\bibitem[Chen \latin{et~al.}(1975)Chen, Silcox, and Vincent]{1975-Chen-Silcox}
Chen,~C.~H.; Silcox,~J.; Vincent,~R. Electron-energy losses in silicon: Bulk and surface plasmons and \ifmmode \check{C}\else \v{C}\fi{}erenkov radiation. \emph{Phys. Rev. B} \textbf{1975}, \emph{12}, 64--71\relax
\mciteBstWouldAddEndPuncttrue
\mciteSetBstMidEndSepPunct{\mcitedefaultmidpunct}
{\mcitedefaultendpunct}{\mcitedefaultseppunct}\relax
\EndOfBibitem
\bibitem[Garc\'{\i}a~de Abajo \latin{et~al.}(2004)Garc\'{\i}a~de Abajo, Rivacoba, Zabala, and Yamamoto]{2004-de-Abajo-Rivacoba}
Garc\'{\i}a~de Abajo,~F.~J.; Rivacoba,~A.; Zabala,~N.; Yamamoto,~N. Boundary effects in Cherenkov radiation. \emph{Phys. Rev. B} \textbf{2004}, \emph{69}, 155420\relax
\mciteBstWouldAddEndPuncttrue
\mciteSetBstMidEndSepPunct{\mcitedefaultmidpunct}
{\mcitedefaultendpunct}{\mcitedefaultseppunct}\relax
\EndOfBibitem
\bibitem[Couillard \latin{et~al.}(2010)Couillard, Yurtsever, and Muller]{2010-Couillard-Yurtsever}
Couillard,~M.; Yurtsever,~A.; Muller,~D.~A. Interference effects on guided Cherenkov emission in silicon from perpendicular, oblique, and parallel boundaries. \emph{Phys. Rev. B} \textbf{2010}, \emph{81}, 195315\relax
\mciteBstWouldAddEndPuncttrue
\mciteSetBstMidEndSepPunct{\mcitedefaultmidpunct}
{\mcitedefaultendpunct}{\mcitedefaultseppunct}\relax
\EndOfBibitem
\bibitem[Saito \latin{et~al.}(2013)Saito, Chen, and Kurata]{2013-Saito-Chen}
Saito,~H.; Chen,~C.~H.; Kurata,~H. Optical guided modes coupled with Čerenkov radiation excited in Si slab using angular-resolved electron energy-loss spectrum. \emph{Journal of Applied Physics} \textbf{2013}, \emph{113}, 113509\relax
\mciteBstWouldAddEndPuncttrue
\mciteSetBstMidEndSepPunct{\mcitedefaultmidpunct}
{\mcitedefaultendpunct}{\mcitedefaultseppunct}\relax
\EndOfBibitem
\bibitem[Preimesberger \latin{et~al.}(2025)Preimesberger, Hornof, Dorfner, Schachinger, Hrto\ifmmode~\check{n}\else \v{n}\fi{}, Kone\ifmmode~\check{c}\else \v{c}\fi{}n\'a, and Haslinger]{2025-Preimesberger-Hornof}
Preimesberger,~A.; Hornof,~D.; Dorfner,~T.; Schachinger,~T.; Hrto\ifmmode~\check{n}\else \v{n}\fi{},~M.; Kone\ifmmode~\check{c}\else \v{c}\fi{}n\'a,~A.; Haslinger,~P. Exploring Single-Photon Recoil on Free Electrons. \emph{Phys. Rev. Lett.} \textbf{2025}, \emph{134}, 096901\relax
\mciteBstWouldAddEndPuncttrue
\mciteSetBstMidEndSepPunct{\mcitedefaultmidpunct}
{\mcitedefaultendpunct}{\mcitedefaultseppunct}\relax
\EndOfBibitem
\bibitem[Karlovets(2011)]{2011-Karlovets}
Karlovets,~D.~V. On the theory of polarization radiation in media with sharp boundaries. \emph{Journal of Experimental and Theoretical Physics} \textbf{2011}, \emph{113}, 27–45\relax
\mciteBstWouldAddEndPuncttrue
\mciteSetBstMidEndSepPunct{\mcitedefaultmidpunct}
{\mcitedefaultendpunct}{\mcitedefaultseppunct}\relax
\EndOfBibitem
\bibitem[Shevelev \latin{et~al.}(2015)Shevelev, Konkov, and Aryshev]{2015-Shevelev-Konkov}
Shevelev,~M.; Konkov,~A.; Aryshev,~A. Soft-x-ray Cherenkov radiation generated by a charged particle moving near a finite-size screen. \emph{Phys. Rev. A} \textbf{2015}, \emph{92}, 053851\relax
\mciteBstWouldAddEndPuncttrue
\mciteSetBstMidEndSepPunct{\mcitedefaultmidpunct}
{\mcitedefaultendpunct}{\mcitedefaultseppunct}\relax
\EndOfBibitem
\bibitem[Inan and Inan(1999)Inan, and Inan]{1999-Inan-Inan}
Inan,~U.~S.; Inan,~A.~S. \emph{Electromagnetic waves}; Prentice Hall, 1999\relax
\mciteBstWouldAddEndPuncttrue
\mciteSetBstMidEndSepPunct{\mcitedefaultmidpunct}
{\mcitedefaultendpunct}{\mcitedefaultseppunct}\relax
\EndOfBibitem
\bibitem[Polyanskiy(2024)]{2024-refractiveindex-info}
Polyanskiy,~M.~N. Refractiveindex.info database of optical constants. \emph{Scientific Data} \textbf{2024}, \emph{11}, 94\relax
\mciteBstWouldAddEndPuncttrue
\mciteSetBstMidEndSepPunct{\mcitedefaultmidpunct}
{\mcitedefaultendpunct}{\mcitedefaultseppunct}\relax
\EndOfBibitem
\bibitem[Saleh and Teich(2019)Saleh, and Teich]{2019-Saleh-Teich}
Saleh,~B.~E.; Teich,~M.~C. \emph{Fundamentals of photonics}; John Wiley \& Sons, 2019\relax
\mciteBstWouldAddEndPuncttrue
\mciteSetBstMidEndSepPunct{\mcitedefaultmidpunct}
{\mcitedefaultendpunct}{\mcitedefaultseppunct}\relax
\EndOfBibitem
\bibitem[Szczepkowicz \latin{et~al.}(2020)Szczepkowicz, Sch\"{a}chter, and England]{2020-Szczepkowicz-Schachter-England}
Szczepkowicz,~A.; Sch\"{a}chter,~L.; England,~R.~J. Frequency-domain calculation of {S}mith{\textendash}{P}urcell radiation for metallic and dielectric gratings. \emph{Applied Optics} \textbf{2020}, \emph{59}, 11146\relax
\mciteBstWouldAddEndPuncttrue
\mciteSetBstMidEndSepPunct{\mcitedefaultmidpunct}
{\mcitedefaultendpunct}{\mcitedefaultseppunct}\relax
\EndOfBibitem
\bibitem[Potylitsyn \latin{et~al.}(2010)Potylitsyn, Ryazanov, Strikhanov, and Tishchenko]{2010-Potylitsyn-Ryazanov}
Potylitsyn,~A.~P.; Ryazanov,~M.~I.; Strikhanov,~M.~N.; Tishchenko,~A.~A. \emph{Diffraction Radiation from Relativistic Particles}; {Springer Tracts in Modern Physics}; Springer Berlin Heidelberg, 2010; Vol. 239\relax
\mciteBstWouldAddEndPuncttrue
\mciteSetBstMidEndSepPunct{\mcitedefaultmidpunct}
{\mcitedefaultendpunct}{\mcitedefaultseppunct}\relax
\EndOfBibitem
\end{mcitethebibliography}

\end{document}